\documentclass[]{imag-ms-template}
\usepackage{textcomp}

\usepackage{graphicx,verbatim}
\usepackage{amssymb}  
\usepackage{pifont}   
\usepackage{color, colortbl}
\usepackage{array} 
\usepackage{multirow}
\usepackage{xcolor} 
\usepackage{tikz}  
\usepackage{soul} 
\usepackage{colortbl}
\usepackage{booktabs}
\usepackage{hyperref}
\usepackage{cleveref}
\usepackage{pifont} 
\usepackage[inkscapelatex=false]{svg}
\usepackage{rotating}
\hypersetup{
    colorlinks=true,
    linkcolor=blue,
    filecolor=cyan,      
    urlcolor=magenta,
    }
\arrayrulecolor{black}
\usepackage[utf8]{inputenc}
\DeclareUnicodeCharacter{2113}{l}

\title{Deep Learning-Based Desikan-Killiany Parcellation of the Brain Using Diffusion MRI} 

\author{Yousef Sadegheih,$^{1}$ Dorit Merhof,$^{1,2\ast}$\\
{\small $^{1}$Faculty of Informatics and Data Science, University of Regensburg, Regensburg, 93053, Germany}\\
{\small $^{2}$Fraunhofer Institute for Digital Medicine MEVIS, Bremen 28359, Germany}\\
{\small $^\ast$Correspondence:  dorit.merhof@ur.de}
}

\begin{document} 

\maketitle

\keywords{Deep Learning, Parcellation and Diffusion MRI}

\begin{abstract}
  Accurate brain parcellation in diffusion MRI (dMRI) space is essential for advanced neuroimaging analyses. However, most existing approaches rely on anatomical MRI for segmentation and inter-modality registration, a process that can introduce errors and limit the versatility of the technique. In this study, we present a novel deep learning-based framework for direct parcellation based on the Desikan-Killiany (DK) atlas using only diffusion MRI data. Our method utilizes a hierarchical, two-stage segmentation network: the first stage performs coarse parcellation into broad brain regions, and the second stage refines the segmentation to delineate more detailed subregions within each coarse category. We conduct an extensive ablation study to evaluate various diffusion-derived parameter maps, identifying an optimal combination of fractional anisotropy, trace, sphericity, and maximum eigenvalue that enhances parellation accuracy. When evaluated on the Human Connectome Project and Consortium for Neuropsychiatric Phenomics datasets, our approach achieves superior Dice Similarity Coefficients compared to existing state-of-the-art models. Additionally, our method demonstrates robust generalization across different image resolutions and acquisition protocols, producing more homogeneous parcellations as measured by the relative standard deviation within regions. This work represents a significant advancement in dMRI-based brain segmentation, providing a precise, reliable, and registration-free solution that is critical for improved structural connectivity and microstructural analyses in both research and clinical applications. The implementation of our method is publicly available on \href{https://github.com/xmindflow/DKParcellationdMRI}{github.com/xmindflow/DKParcellationdMRI}.
\end{abstract}

\section{Introduction}

Diffusion Magnetic Resonance Imaging (dMRI) is a valuable imaging technique that captures the movement of water molecules within biological tissues, offering indirect yet detailed information about microstructural organization~\citep{basser2011microstructural}. Unlike traditional anatomical MRI, which focuses on tissue contrast and morphology, dMRI reveals the structure and connectivity of white matter pathways through fiber tracking and tractography~\citep{basser2002diffusion}. As a result, dMRI has become widely used in both research and clinical settings for investigating brain development, aging, and neurological conditions~\citep{zhang2022quantitative}.

An essential step in analyzing brain imaging data is parcellation, which divides the brain into meaningful anatomical or functional regions. This step provides a framework for regional analysis and supports tasks such as structural connectivity mapping and disease-related abnormality detection~\citep{cloutman2012connectivity, ji2019mapping, wassermann2016white, sporns2005human, seitz2018alteration}. In diffusion MRI studies, accurate parcellation is especially important for building reliable connectomes and assessing localized tissue changes~\citep{zhang2023ddparcel}.

Traditionally, brain parcellation is performed on high-resolution anatomical MRI (typically T1- or T2-weighted scans) using established software tools such as FreeSurfer~\citep{fischl2012freesurfer} or CAT12~\citep{gaser2024cat}. The anatomical parcellations are then aligned with the diffusion images through inter-modality registration. However, this process is error-prone due to distortions introduced by echo-planar imaging (EPI) used in dMRI~\citep{wu2008comparison, jones2010twenty, albi2018image} and the lower spatial resolution of diffusion data~\citep{malinsky2013registration}. Misregistration between modalities can lead to inaccurate label assignments and compromise downstream analyses. Furthermore, anatomical MRI may not always be available in certain clinical or research workflows~\citep{zhang2023ddparcel}.

To address these issues, recent studies have developed deep learning methods that perform brain parcellation directly on dMRI data~\citep{liu2007brain, wen2013brain, yap2015brain, ciritsis2018automated, zhang2021deep, theaud2022doris, zhang2023ddparcel, li2025ddevenet}. These approaches often use convolutional neural networks (CNNs) to extract features from diffusion-derived maps such as fractional anisotropy (FA) and mean diffusivity (MD). Advanced methods like DDParcel~\citep{zhang2023ddparcel} and DDEvENet~\citep{li2025ddevenet} have achieved notable performance by combining multiple input features and ensemble learning to improve segmentation quality without relying on anatomical MRI.

However, current dMRI-based parcellation methods still face several challenges. Many approaches rely on 2D or 2.5D representations, where networks are trained on individual slices in axial, coronal, or sagittal views. Although this design, seen in FastSurfer~\citep{henschel2020fastsurfer}, reduces computational load, it limits the ability to fully capture 3D spatial relationships in the brain. This can result in reduced segmentation accuracy, especially in regions with complex anatomy~\citep{chen2024xlstm}. Fully 3D models have shown improved results in medical image segmentation tasks by learning volumetric features and context~\citep{isensee2021nnu, azad2024beyond}, which can be beneficial for detailed brain parcellation.

Another challenge lies in the choice of input features for parcellation. Current models use different combinations of diffusion parameters, but there is no clear agreement on which combination offers the best performance. For example, DDParcel~\citep{zhang2023ddparcel} uses Fractional Anisotropy (FA), Mean Diffusivity (MD), and tensor eigenvalues in a multi-branch network, while DDEvENet~\citep{li2025ddevenet} employs a confidence-weighted ensemble approach with different tensor eigenvalues. This lack of standardization in feature selection can affect generalization across datasets with varying scanning protocols~\citep{sadegheih2024segmentation}.

Finally, segmentation models based on the Desikan-Killiany (DK) atlas~\citep{desikan2006automated} face challenges related to imbalanced label distributions. The DK atlas~\citep{desikan2006automated} is a widely used brain parcellation scheme in neuroimaging, particularly for structural and connectivity analyses. It divides the cerebral cortex into 101 regions (including both hemispheres), based on anatomical landmarks derived from gyral and sulcal patterns. These regions vary substantially in size and volume across the brain, which presents a challenge during model training. Larger brain regions tend to dominate the loss function during model training, while smaller regions are underrepresented, leading to poorer segmentation of these finer anatomical structures~\citep{sadegheih2024segmentation}. This imbalance can hinder tasks that require precise boundary definitions in all brain regions, making it a significant issue for accurate parcellation.

In this paper, we present a 3D deep learning framework designed for direct DK parcellation using only diffusion MRI. Our method is based on a hierarchical, coarse-to-fine segmentation strategy that first predicts larger anatomical divisions and then refines them into detailed parcels. By using fully 3D networks, our model can better learn spatial patterns and boundaries from volumetric data. Additionally, we perform a thorough comparison of diffusion-derived parameters to identify those that contribute most to segmentation performance.

Our main contributions are: \ding{182} A detailed ablation study on the influence of different diffusion-derived features on parcellation accuracy. \ding{183} A hierarchical 3D segmentation architecture that improves performance in both large and small brain regions. \ding{184} Demonstrated improvement over state-of-the-art dMRI parcellation methods, especially in low-resolution scenarios.

Through this work, we aim to provide a reliable and practical solution for brain parcellation directly in the diffusion space, removing the need for anatomical MRI and complex registration steps. This contributes to more efficient and accessible analysis of diffusion MRI data in both research and clinical contexts.

Throughout this paper, numerous acronyms are used. The reader is encouraged to refer to \Cref{tab:notations} for the full list and definitions of these acronyms.  

\begin{table}[!ht]
    \centering
    \setlength{\tabcolsep}{5pt}
    \caption{
    Key notations and acronyms used throughout the paper, organized by category and color-coded for clarity. Green represents generic acronyms, magenta is used for metrics, and orange highlights derived parameters.
    }
    \begin{tabular}{r|p{8cm}}
    \bottomrule
    \rowcolor{gray!10}
    \textbf{Acronym} & \textbf{Meaning} \\ 
    \hline

    \rowcolor{green!10}
    $\textit{MRI}$ & Magnetic Resonance Imaging\\    
    \rowcolor{green!5}
     $\textit{dMRI}$ & Diffusion MRI\\
    \rowcolor{green!10}
    $\textit{DWI}$  & Diffusion-Weighted Imaging\\
    \rowcolor{green!5}
    $\textit{DTI}$ & Diffusion Tensor Imaging\\
    \rowcolor{green!10}
    $\textit{DK}$ & Desikan-Killian\\
    \rowcolor{green!5}
    $\textit{CNN}$ & Convolutional Neural Network\\
    \rowcolor{green!10}
    $\textit{CNP}$ & Consortium for Neuropsychiatric Phenomics \\
    \rowcolor{green!5}
    $\textit{HCP}$ & Human Connectome Project\\
    \rowcolor{green!10}
    $\textit{ADHD}$ & attention-deficit/hyperactivity disorder \\
    \rowcolor{green!5}
    $\textit{MNI}$ & Montreal Neurological Institute\\
    \rowcolor{green!10}
    $\textit{EPI}$  & Echo Planar Imaging\\    
    \rowcolor{green!5}
    $\textit{ANT}$ & Advanced Normalization Tools\\
    \rowcolor{green!10}
    $\textit{FS}$ & FreeSurfer\\    
    \rowcolor{green!5}
    $\textit{LIA}$ & Left-to-right, Inferior-to-superior, Anterior-to-posterior orientation\\

    \rowcolor{magenta!10}
    $\textit{STD}$ &  Standard Deviation\\
    \rowcolor{magenta!5}
    $\textit{CV}$ &  Coefficient of Variation\\
    \rowcolor{magenta!10}
    $\textit{RSD}$ & Relative Standard Deviation\\
    \rowcolor{magenta!5}
    $\textit{HD95}$ & 95th percentile Hausdorff Distance\\
    \rowcolor{magenta!10}
    $\textit{DSC}$ & Dice Similarity Coefficient \\

    \rowcolor{orange!5}
    $\textit{F/FA}$ & Fractional Anisotrop\\
    \rowcolor{orange!10}
    $\textit{T}$ & Trace (the sum of the eigenvalues) \\   
    \rowcolor{orange!5}
    $\textit{MD}$ & Mean Diffusivity \\
    \rowcolor{orange!10}
    $\textit{S/CS}$ & Sphericity\\
    \rowcolor{orange!5}
    $\textit{L/CL}$ & Linearity\\
    \rowcolor{orange!10}
    $\textit{P/CP}$ & Planarity\\
    \rowcolor{orange!5}
    $\textit{E1}$ & Maximum Eigenvalue\\
    \rowcolor{orange!10}
    $\textit{E2}$ & Mid Eigenvalue\\
    \rowcolor{orange!5}
    $\textit{E3}$ & Minimum Eigenvalue\\

    \toprule
    \end{tabular}
\label{tab:notations}
\end{table}
\section{Datasets and preporcess}
\subsection{Datasets} 
Our evaluation utilized two datasets: the Human Connectome Project (HCP)~\citep{van2013wu} and the Consortium for Neuropsychiatric Phenomics (CNP)~\citep{poldrack2016phenome}. For the HCP dataset, we used a subset of 100 young, healthy adults, consisting of 46 males and 54 females, with an average age of 29.1 years. The dataset was split into training, validation, and test sets with a ratio of 50:30:20, respectively. 

The dMRI data from the HCP dataset have a voxel size of $1.25^3 mm^3$ and include 18 baseline images along with 270 diffusion-weighted images from three b-shells, with b-values of 1000, 2000, and 3000 $s/mm^2$. For compatibility with legacy data structures, we only used the b = 1000 $s/mm^2$ shell in our analysis. The structural MRI data (T1w and T2w) for this dataset has a voxel size of $0.7^3 mm^3$.

The CNP dataset consists of 272 young adults, including individuals with various health conditions such as schizophrenia, bipolar disorder, and ADHD, as well as healthy controls. However, only 214 participants had the necessary structural and dMRI data for our analysis. The cohort included 114 males and 100 females, with an average age of 33.2 years. Of these, 108 were healthy controls, 41 had bipolar disorder, 36 had ADHD, and 29 had schizophrenia. Each dMRI scan in the CNP dataset has a voxel size of $2^3mm^3$ and contains one baseline image along with 64 diffusion-weighted images with a b-value of 1000 $s/mm^2$. The structural MRI (T1w) data has a voxel size of $1^3mm^3$.

\subsection{Preprocessing} \label{sub:preprocessing}
We apply different preprocessing steps for the training and inference phases.

\begin{figure}[!t]
    \centering
    \includegraphics[width=0.85\textwidth]{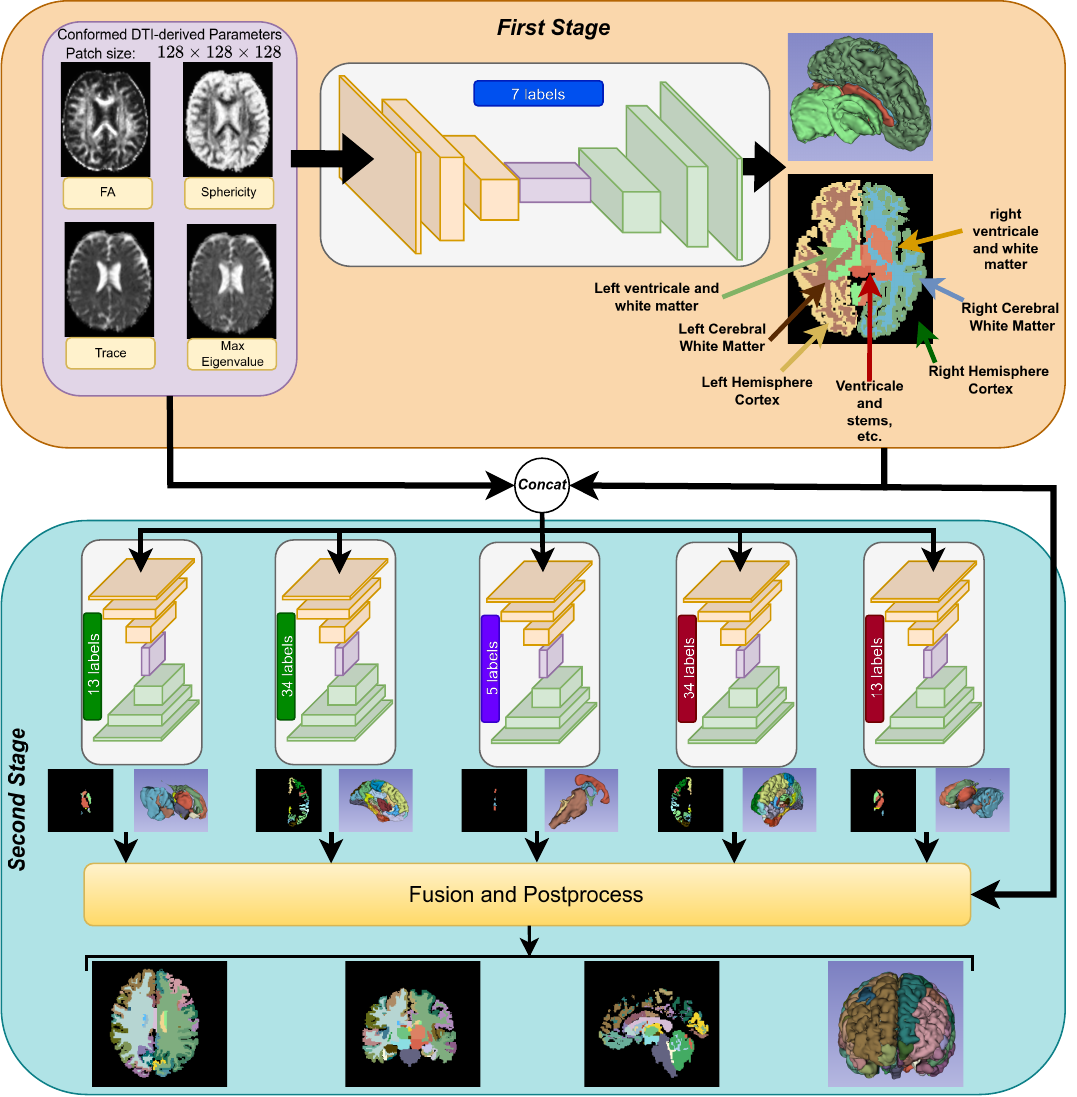}
    \caption{Hierarchical two-stage framework overview.}
    \label{fig:framework}
\end{figure}

For training, we use data from the HCP dataset, which has demonstrated high reliability as ground truth. The HCP data is preprocessed using the HCP minimal processing pipeline~\citep{glasser2013minimal}. This pipeline includes motion correction, eddy current correction, EPI distortion correction, and registration to the 6th generation nonlinear MNI152 space~\citep{grabner2006symmetric}. A brain mask is also extracted during this process. Prior studies have shown that FreeSurfer parcellation obtained from anatomical MRI, when coregistered to dMRI space, can serve as a reliable ground truth for parcellation tasks~\citep{zhang2021deep,zhang2023ddparcel}.

We use the white matter parcellation generated by FreeSurfer, which we convert to the DK atlas by merging some labels. This approach is justified because the HCP pipeline performs extensive postprocessing on white matter parcellation to ensure accurate identification of anatomical regions. Having a well-coregistered parcellation in dMRI space allows us to perform parcellation similarly to a typical segmentation task. Since the dMRI data contains three b-value shells, we exclude the b=2000 and b=3000 shells to focus on the parameters derivable from the remaining data.

For the CNP data, the dMRI data is not preprocessed in the same way as HCP data. Therefore, we apply a validated preprocessing pipeline~\citep{tashrif_billah_2020_4118796} that includes eddy current correction and motion correction using FSL tools~\citep{jenkinson2012fsl}, along with brain masking extraction~\citep{cetin2024harmonized,Palanivelu_CNN_based_diffusion_2024}. Since T2-weighted images are not available, EPI distortion correction is omitted. Following established protocols~\citep{zhang2023ddparcel,cetin2024harmonized,di2021white}, we obtain parcellation in the dMRI baseline space by nonlinearly coregistering the FreeSurfer parcellation to the dMRI space using ANTs~\citep{avants2009advanced}. The parcellation is then resampled to dMRI space using nearest neighbor interpolation. This coregistration and parcellation serve as the baseline reference for the parcellation task.

After preprocessing the DWI data, we extract the necessary diffusion parameters. These include FA, trace, sphericity~\citep{westin2002processing}, and the maximum eigenvalue maps (For a detailed discussion on the rationale behind the selection of these parameters, refer to \Cref{sub:parameter_impact}). We use 3DSlicer~\citep{fedorov20123d} with the SlicerDMRI extension~\citep{norton2017slicerdmri} to fit the diffusion tensor model from the DWI data using a least squares method. Then, the required parameter maps are calculated using SlicerDMRI.

To account for head motion during MRI acquisition, all derived parameter maps are transformed to the nonlinear 6th generation MNI152 space~\citep{grabner2006symmetric}. Since FreeSurfer performs parcellation in a standardized conformed space, we apply this conformation to the parameter maps as well. This ensures all maps are resampled to a uniform spacing of $1\times1\times1\ \mathrm{mm}^3$, a resolution of $256^3$, and LIA orientation, which matches FreeSurfer's output.

Throughout the pipeline, all data and derived parameters remain co-registered in this conformed MNI space. After parcellation prediction, the results are transformed back to the original native space to align with the original DWI data. We perform these transformations using the baseline (b=0) image as the reference. First, the transformation from native space to MNI space is calculated. Then, the derived parameters are moved to MNI space using this transformation. Finally, the inverse transform is applied to return the parcellation results to native space.

Our model training uses the nnUNet framework, which applies the same preprocessing steps for all MRI-based inputs, including the derived diffusion parameter maps.

\section{Methodolodgy}


The proposed training strategy enhances segmentation accuracy by leveraging the hierarchical structure of brain parcellation. It consists of two stages: first, a coarse segmentation (\Cref{sub:coarse_parcel}) partitions the brain into broad regions, and then, a refinement step within each region (\Cref{sub:fine_parcel}) enables the delineation of finer anatomical parcels. This two-stage approach ensures precise label assignment to dMRI data, with each stage optimized for its specific task to improve overall segmentation quality.

It is important to highlight that our methodology includes a preprocessing step, which was detailed in Section \ref{sub:preprocessing}. In this section, we focus on the processing and post-processing (\Cref{sub:post_process}) procedures. A visual representation of the overall workflow is provided in Figure \ref{fig:framework}.

While the framework is compatible with any 3D model, the selection of the model plays a crucial role in determining the final performance. Stronger, more robust segmentation models generally yield better results. 

\subsection{Coarse Parcellation}
\label{sub:coarse_parcel}

In the coarse parcellation stage, a single 3D model is used for segmentation; however, it does not directly predict all 101 labels of the DK atlas. Instead, the brain regions are grouped into seven broader categories to simplify the segmentation process: left and right cerebral white matter, left and right cortical regions, central areas (including the corpus callosum, cerebrospinal fluid, brainstem, and ventricles), and the remaining left and right regions.

The model input consists of 3D patches derived from four dMRI-based parameter maps: fractional anisotropy, trace, sphericity~\citep{westin2002processing}, and maximum eigenvalue. These maps are spatially aligned and resampled to a consistent isotropic resolution of 1mm³, conforming to the LIA orientation and a volume size of 256³. The model’s output is a segmentation map consisting of the seven predefined categories.

This grouping strategy is intended to balance the size of the regions, allowing the model to focus first on larger anatomical areas. Notably, the cerebral white matter regions, which account for the majority of brain volume, are segmented in this stage and are directly assigned two of the final 101 labels without further subdivision. This approach enables the model to concentrate on more complex, smaller regions, such as cortical structures, in the subsequent refinement stage.

\subsection{Fine Parcellation (Sub-region Segmentation)}
\label{sub:fine_parcel}

In the fine parcellation stage, the remaining 99 regions are segmented, excluding the left and right cerebral white matter regions that were defined in the coarse parcellation stage. To achieve this, we employ five distinct segmentation networks, each dedicated to a specific category from the coarse segmentation: left cortical regions, right cortical regions, central areas, and the remaining regions for both hemispheres.

Each network receives as input the same four dMRI-derived parameter maps (\Cref{sub:coarse_parcel}) and the coarse segmentation masks. These inputs are combined into a five-channel volume, where four channels correspond to the dMRI parameter maps and one channel represents the coarse segmentation mask. This mask is normalized to a range between 0 and 1, ensuring that label values from the coarse segmentation do not disproportionately influence the input scale.

As illustrated in Figure \ref{fig:framework}, these five networks produce segmented regions corresponding to cortical areas (divided into 34 labels), central regions (five labels), and the remaining regions (14 labels per hemisphere). The outputs from these independent networks are then combined to generate the complete fine parcellation, which provides a detailed segmentation of the brain into 99 sub-regions.

\subsection{Post-processing}
\label{sub:post_process}

Following the fine segmentation, all generated labels are merged to form the complete set of 101 parcellation labels. These labels are then transformed back from the conformed space (used during preprocessing) to the original dMRI space, utilizing the inverse transformation applied during preprocessing.

In the post-processing step, the coarse segmentation serves as a mask, restricting the fine segmentations to their corresponding coarse regions. This ensures that each subregion is confined within the boundaries defined by the coarse parcellation. Any subregion that falls outside the coarse segmentation mask is removed. A dilation step is then applied to the remaining fine segmentations to smooth out the boundaries, improving the continuity and accuracy of the regions.

To further refine the segmentation, we apply a largest connected component analysis to remove small, isolated regions that may arise from segmentation artifacts or noise. Given the expectation that brain regions should be contiguous, this step retains only the largest connected component for each label, ensuring that small, disconnected regions are excluded from the final parcellation.

Subsequently, the combined labels are resampled to match the original dMRI resolution using nearest neighbor interpolation, implemented in 3DSlicer~\citep{fedorov20123d}. This resampling produces the final parcellation, which is then aligned with the original dMRI data. Finally, the labels are mapped to the corresponding FreeSurfer lookup table labels.



\section{Experimental setup and metrics}
\subsection{Experimental setup}
The pipeline and framework was implemented using modules from the nnUNet framework~\citep{isensee2021nnu} while using PyTorch 2.1.0~\citep{paszke2019pytorch}. Each stage of the framework was trained with a batch size of 2 and a patch size of $128^3$. For training, each stage used a single NVIDIA A100 GPU (80 GB VRAM). The learning rate settings followed the original hyperparameters outlined in the related work. Specifically, for the U-Net~\citep{isensee2021nnu} architecture, we used the SGD optimizer with Nesterov momentum (0.99) and a weight decay of \(3\times 10^{-5}\). The learning rate followed a polynomial decay strategy (power 0.9) with an initial value of 0.01. For the MedNeXt~\citep{roy2023mednext} model, we used the AdamW optimizer~\citep{loshchilov2017decoupled} with an initial learning rate of 0.001 and applied the same polynomial decay strategy. For SwinUNETR~\citep{hatamizadeh2021swin}, we have used the same learning rate as the nnUNet.

The training process used a composite loss function combining DICE~\citep{milletari2016v} and cross-entropy, weighted equally (1:1). Data augmentation techniques followed those described in~\citep{zhou2021nnformer,shaker2024unetr++,isensee2021nnu}. Training was carried out for 1000 epochs, with 250 iterations per epoch, leading to a total of 250,000 iterations.

To generate the final prediction for each stage, we used a 0.5 overlap along each spatial dimension and applied a Gaussian kernel to reduce the weight of predictions made at the edges of the windows. The training time varied depending on the number of labels predicted at each stage and the specific model used. On average, the U-Net version required 25 GPU hours, the MedNeXt model took an average of 40 GPU hours, and the SwinUNETR model took around 32 GPU hours.

\subsection{Evaluation Metric}

For evaluation, we use the DSC and the HD95 on the HCP test data, where ground truth parcellations are available.

For the CNP data, since no ground truth is provided, we assess the quality of parcellation by calculating the relative standard deviation (RSD), also known as the coefficient of variation (CV), for the FA, MD, and sphericity values within each parcellated region. The RSD for each region is defined as:

\[
RSD = \frac{\mathrm{std}}{\mathrm{mean}}
\]

Previous studies in diffusion MRI~\citep{zhang2023ddparcel,roberts2017consistency,zhang2017comparison} have shown that lower RSD values indicate greater homogeneity within regions, which reflects better parcellation quality.

\section{Results and Discussion}

\begin{table*}[t]
\renewcommand{\r}[1]{\textcolor{red}{#1}}
\renewcommand{\b}[1]{\textcolor{blue}{#1}}
\centering
\caption{Diffusion-derived parameter ablation study using 2D nnUNet (F: FA, T: Trace, S: Sphericity, P: Planarity, L: Linearity, E1: Max eigenvalue, E2: Mid eigenvalue, E3: Min eigenvalue). \b{Blue} and \r{red} indicate the best and second-best results in each modality section, respectively. Results are reported as DSC with standard deviation in parentheses.}
\label{tab:biomarker_study}
\resizebox{\textwidth}{!}{%
\begin{tabular}{c|cc|cccc|cc|cccc|c|c}
\toprule
\rowcolor{gray!10}  
                            & \multicolumn{2}{c|}{Param.}                 & \multicolumn{2}{c|}{DSC}                                              & \multicolumn{2}{c|}{Param.}                  & \multicolumn{2}{c|}{DSC}                  & \multicolumn{2}{c|}{Param.}                                              & \multicolumn{2}{c|}{DSC}                 & Param.            & DSC          \\ \hline
\multirow{2}{*}{1 Modality} & \multicolumn{2}{c|}{F}                   & \multicolumn{2}{c|}{74.15 (0.05)}                                     & \multicolumn{2}{c|}{T}                    & \multicolumn{2}{c|}{73.16 (0.01)}         & \multicolumn{2}{c|}{L}                                                & \multicolumn{2}{c|}{71.93 (0.05)}        & P              & 69.97 (0.08) \\ \cline{2-15} 
                            & \multicolumn{2}{c|}{S}                   & \multicolumn{2}{c|}{74.34 (0.08)}                                     & \multicolumn{2}{c|}{E1}                   & \multicolumn{2}{c|}{\r{74.39 (0.06)}}         & \multicolumn{2}{c|}{E2}                                               & \multicolumn{2}{c|}{73.51 (0.12)}        & E3             & \b{75.01 (0.09)} \\ \midrule
                            \rowcolor{gray!10} 
                            & \multicolumn{1}{c|}{Param.}  & DSC          & \multicolumn{1}{c|}{Param.}          & \multicolumn{1}{c|}{DSC}          & \multicolumn{1}{c|}{Param.}   & DSC          & \multicolumn{1}{c|}{Param.}   & DSC          & \multicolumn{1}{c|}{Param.}          & \multicolumn{1}{c|}{DSC}          & \multicolumn{1}{c|}{Param.}  & DSC          & Param.            & DSC          \\ \hline
\multirow{3}{*}{2 Modality} & \multicolumn{1}{c|}{F+T}  & \b{76.34 (0.05)} & \multicolumn{1}{c|}{F+L}          & \multicolumn{1}{c|}{74.37 (0.09)} & \multicolumn{1}{c|}{F+P}   & 74.44 (0.08) & \multicolumn{1}{c|}{F+S}   & 74.72 (0.17) & \multicolumn{1}{c|}{T+L}          & \multicolumn{1}{c|}{75.40 (0.13)} & \multicolumn{1}{c|}{T+P}  & 74.89 (0.12) & T+S            & 76.26 (0.07) \\ \cline{2-15} 
                            & \multicolumn{1}{c|}{L+P}  & 74.45 (0.06) & \multicolumn{1}{c|}{L+S}          & \multicolumn{1}{c|}{74.69 (0.09)} & \multicolumn{1}{c|}{E1+E2} & 75.52 (0.05) & \multicolumn{1}{c|}{E1+E3} & 76.27 (0.04) & \multicolumn{1}{c|}{E2+E3}        & \multicolumn{1}{c|}{75.70 (0.07)} & \multicolumn{1}{c|}{F+E1} & 76.28 (0.10) & F+E2           & \r{76.33 (0.07)} \\ \cline{2-15} 
                            & \multicolumn{1}{c|}{F+E3} & 76.28 (0.15) & \multicolumn{1}{c|}{T+E1}         & \multicolumn{1}{c|}{75.90 (0.03)} & \multicolumn{1}{c|}{T+E2}  & 75.16 (0.06) & \multicolumn{1}{c|}{T+E3}  & 76.12 (0.09) & \multicolumn{1}{c|}{S+E1}         & \multicolumn{1}{c|}{76.32 (0.02)} & \multicolumn{1}{c|}{S+E2} & 76.26 (0.08) & S+E3           & 76.29 (0.07) \\ \midrule
                            \rowcolor{gray!10} 
                            & \multicolumn{2}{c|}{Param.}                 & \multicolumn{1}{c|}{DSC}          & \multicolumn{2}{c|}{Param.}                                       & DSC          & \multicolumn{2}{c|}{Param.}                  & \multicolumn{1}{c|}{DSC}          & \multicolumn{2}{c|}{Param.}                                      & DSC          & Param.            & DSC          \\ \hline
\multirow{2}{*}{3 Modality} & \multicolumn{2}{c|}{E1+E2+E3}            & \multicolumn{1}{c|}{76.28 (0.05)} & \multicolumn{2}{c|}{F+E1+E3}                                   & 76.35 (0.14) & \multicolumn{2}{c|}{T+E1+E3}              & \multicolumn{1}{c|}{76.15 (0.08)} & \multicolumn{2}{c|}{S+E1+E3}                                  & 76.32 (0.20) & T+S+E3         & 76.23 (0.11) \\ \cline{2-15} 
                            & \multicolumn{2}{c|}{T+S+E1}              & \multicolumn{1}{c|}{76.36 (0.07)} & \multicolumn{2}{c|}{T+S+F}                                     & \b{76.48 (0.10)} & \multicolumn{2}{c|}{T+F+E1}               & \multicolumn{1}{c|}{\r{76.41 (0.02)}} & \multicolumn{2}{c|}{T+F+E3}                                   & 76.37 (0.09) &                &              \\ \midrule
4 Modality                  & \multicolumn{2}{c|}{F+T+E2+E3}           & \multicolumn{1}{c|}{76.39 (0.15)} & \multicolumn{2}{c|}{T+S+E1+E3}                                 & 76.16 (0.15) & \multicolumn{2}{c|}{T+F+S+E3}             & \multicolumn{1}{c|}{\r{76.49 (0.10)}} & \multicolumn{2}{c|}{T+F+S+E1}                                 & \b{76.52 (0.08)} & T+F+E1+E3      & 76.46 (0.09) \\ \midrule
\rowcolor{gray!10} 
                            & \multicolumn{2}{c|}{Param.}                 & \multicolumn{2}{c|}{DSC}                                              & \multicolumn{2}{c|}{Param.}                  & \multicolumn{2}{c|}{DSC}                  & \multicolumn{2}{c|}{Param.}                                              & \multicolumn{2}{c|}{DSC}                 & Param.            & DSC          \\ \hline
\textgreater{}4 Modality    & \multicolumn{2}{c|}{T+F+E1+E2+E3}        & \multicolumn{2}{c|}{76.42 (0.07)}                                     & \multicolumn{2}{c|}{T+S+E1+E2+E3}         & \multicolumn{2}{c|}{76.37 (0.06)}         & \multicolumn{2}{c|}{T+F+S+E1+E3}                                      & \multicolumn{2}{c|}{\b{76.47 (0.06)}}        & T+F+S+E1+E2+E3 & \r{76.41 (0.07)} \\ \bottomrule
\end{tabular}

}
\end{table*}

\subsection{Parameter Impact Study} \label{sub:parameter_impact}

\Cref{tab:biomarker_study} presents the results of the parameter impact study, which investigates how different parameters derived from DTI data contribute to the DK parcellation task when using a deep learning approach. The motivation for this analysis stems from the variety of parameters employed in existing literature~\citep{zhang2023ddparcel, li2025ddevenet} and the recognition that a thorough modality impact study is critical for deep learning-based segmentation tasks. As some parameters may overlap or introduce redundancy into the network, understanding the individual and combined effects of these parameters is essential for optimizing model performance~\citep{sadegheih2024segmentation}.

For this study, we selected the 2D nnUNet architecture~\citep{isensee2021nnu} due to its efficient training time and ability to provide a fair baseline for comparing different parameter inputs. The parameters used in our modality impact study include the eigenvalues from the DTI data: maximum eigenvalue, mid eigenvalue, and minimum eigenvalue. From these eigenvalues, several derived parameters can be calculated, such as Trace (the sum of the eigenvalues), Fractional Anisotropy (FA), Linearity (CL), Planarity (CP), and Sphericity (CS). Specifically, FA is computed using the formula: 

\[
FA = \sqrt{\frac{(\lambda_1 - \lambda_2)^2 + (\lambda_2 - \lambda_3)^2 + (\lambda_1 - \lambda_3)^2}{2(\lambda_1^2 + \lambda_2^2 + \lambda_3^2)}}
\]

and the other parameters are similarly derived from the eigenvalues: 

\[
CL = \frac{\lambda_1 - \lambda_2}{\lambda_1 + \lambda_2 + \lambda_3}, \quad CP = \frac{2(\lambda_2 - \lambda_3)}{\lambda_1 + \lambda_2 + \lambda_3}, \quad CS = \frac{3\lambda_3}{\lambda_1 + \lambda_2 + \lambda_3}
\]

Our results in~\Cref{tab:biomarker_study} indicate that diffusion-derived parameters each provide distinct contributions to cortical and subcortical parcellation when individually used as input to a convolutional neural network. The minimum eigenvalue (E3) stands out by producing the highest DSC among all single-parameter inputs. This outcome suggests that E3 is particularly sensitive to microstructural differences that are vital for accurately separating anatomical regions. This can be explained by E3’s close relationship with radial diffusivity, reflecting the degree of restriction experienced by water molecules as they move perpendicular to the dominant orientation of cellular structures. Such sensitivity is especially helpful in the cortex and subcortex, where subtle differences in the arrangement and packing of cells define many boundaries. A likely reason for E3’s superiority over E1 is its higher specificity for non-white matter regions. E1, which reflects axial diffusivity, is often elevated in both white matter and cerebrospinal fluid, thereby reducing its ability to pinpoint boundaries between gray and white matter or between brain tissue and fluid-filled spaces. In contrast, E3 tends to be lower in restricted tissues and higher in more isotropic environments, allowing it to better highlight regional changes in tissue properties. As a result, E3 provides the neural network with clearer signals for distinguishing both cortical microstructure and regions near the ventricles.
Shape-based parameters such as CL and CP show limited effectiveness when used alone. Their relatively low performance suggests that while these measures reflect certain geometric aspects of water diffusion, they do not offer sufficient contrast or specificity to allow reliable identification of anatomical boundaries. Interestingly, the sphericity parameter (CS), which measures how close diffusion is to being the same in all directions, performs better than other shape-based metrics. This is notable, as sphericity is not typically emphasized in previous studies, but our results indicate that it can add meaningful information about tissue organization, possibly by highlighting areas with more isotropic diffusion, such as certain gray matter regions or transition zones at tissue interfaces. Notably, E3 outperforms other common metrics such as FA, Trace, CS, and even E1. This suggests that the raw eigenvalues, especially E3, capture more detailed information about the tissue than composite metrics, which average or combine different features into one value.

When examining dual-parameter combinations, we observe further differences in performance and synergy. Although Trace alone yields only moderate results, its combination with FA achieves the highest DSC among all two-parameter inputs. This finding indicates that Trace and FA capture distinct, yet complementary, aspects of tissue structure: Trace measures the total amount of diffusion, while FA quantifies how directional the diffusion is. The network thus benefits from receiving information about both the strength and the pattern of diffusion, allowing for more precise parcellation.

Additionally, combining Trace with CS leads to improved results that are comparable to those achieved with the Trace-FA pair. This can be understood by examining the relationship between the shape metrics FA and S. Both FA and S measure how diffusion deviates from isotropy, but in opposite ways: FA is high when diffusion is strongly directional, while S is high when diffusion is nearly equal in all directions. Our analysis reveals a strong negative correlation between FA and S, meaning they largely provide overlapping information. This is confirmed by our findings that pairing FA and S without Trace does not improve model performance. In some cases, such combinations may even be counterproductive, as overlapping features reinforce the same signals instead of adding new insight. However, when either shape metric is combined with Trace, the network benefits from having access to both shape and overall diffusion magnitude, which results in better parcellation. These results highlight the importance of selecting input parameters that are not only informative but also non-redundant, as combining measures that capture different aspects of tissue properties is key for effective parcellation.



\begin{table*}[!t]
\renewcommand{\r}[1]{\textcolor{red}{#1}}
\renewcommand{\b}[1]{\textcolor{blue}{#1}}
\centering
\caption{Results of the test set of the HCP dataset compared to the SOTA architectures. \b{Blue} and \r{red} indicate the best and second-best results, respectively. Values are reported as the mean with the standard deviation in parentheses.}
\label{tab:SOTA}
\resizebox{\textwidth}{!}{%
\begin{tabular}{c||c|c|c}
\bottomrule
\rowcolor{gray!10}
Dimension                         & Model                                                           & DSC  \(\uparrow\)                             & HD95 \(\downarrow\)                              \\ \toprule
\multirow{3}{*}{\begin{sideways}2.5 D\end{sideways}}       & FastSurfer~\citep{henschel2020fastsurfer} & 75.57  (0.14)                      & 2.571  (0.008)                      \\ \cline{2-4}
                             & DDParcel~\citep{zhang2023ddparcel}        & 78.96  (0.12)                      & 1.682  (0.011)                      \\ \cline{2-4}
                             & DDEvENet~\citep{li2025ddevenet}           & 78.55  (0.12)                      & 1.684  (0.011)                      \\ \midrule
\multirow{11}{*}{\begin{sideways}General 3D\end{sideways}} & nnUNet~\citep{isensee2021nnu}             & 79.64  (0.09)                      & 1.523  (0.007)                      \\ \cline{2-4}
                             & SwinUNETR~\citep{hatamizadeh2021swin}     & 79.05  (0.03)                      & 1.575  (0.004)                      \\ \cline{2-4}
                             & MedNeXt-M-K3~\citep{roy2023mednext}       & \r{81.35  (0.10)} & \r{1.474  (0.010)} \\ \cline{2-4}
                             & LHU-Net~\citep{sadegheih2024lhu}          & 74.66  (0.03)                      & 1.903  (0.010)                      \\ \cline{2-4}
                             & UNETR~\citep{hatamizadeh2022unetr}        & 76.17  (0.06)                      & 1.752  (0.020)                      \\ \cline{2-4}
                             & UNETR++~\citep{shaker2024unetr++}         & 75.80  (0.03)                      & 1.778  (0.018)                      \\ \cline{2-4}
                             & CoTR~\citep{xie2021cotr}                  & 77.60  (0.11)                      & 1.608  (0.010)                      \\ \cline{2-4}
                             & nnFormer~\citep{zhou2021nnformer}         & 74.60  (0.10)                      & 1.800  (0.011)                      \\ \cline{2-4}
                             & SegFormer3D~\citep{perera2024segformer3d} & 71.72  (0.56)                      & 2.082  (0.080)                      \\ \cline{2-4}
                             & SwinUNETR-V2~\citep{he2023swinunetr} &  78.85  (0.12)   &  1.579  (0.011) \\ \cline{2-4}
                             & TransBTS~\citep{wenxuan2021transbts}      & 76.43  (0.04)                      & 1.667  (0.038)                      \\ \bottomrule
\cellcolor[HTML]{C8FFFD}     & \cellcolor[HTML]{C8FFFD}$OURS_{unet}$                                                   &\cellcolor[HTML]{C8FFFD} 81.12  (0.06)                      & \cellcolor[HTML]{C8FFFD} \b{1.469  (0.009)} \\ \cline{2-4}

 \cellcolor[HTML]{C8FFFD}    & \cellcolor[HTML]{C8FFFD}$OURS_{MedNeXt}$                                                & \cellcolor[HTML]{C8FFFD}\b{82.09  (0.05)} & \cellcolor[HTML]{C8FFFD}\r{1.474  (0.011)} \\ \cline{2-4}

\multirow{-3}{*}{\cellcolor[HTML]{C8FFFD}\begin{sideways}Ours 3D\end{sideways}}                             & \cellcolor[HTML]{C8FFFD}$OURS_{SwinUNETR}$                                              & \cellcolor[HTML]{C8FFFD}79.92  (0.03)                      & \cellcolor[HTML]{C8FFFD}1.573  (0.005)  \\
                             \toprule
\end{tabular}

}
\end{table*}

The analysis of three-parameter combinations further emphasizes the value of complementary information. The set of Trace, S, and FA results in a significant performance boost, suggesting that including both magnitude (Trace) and shape (FA, S) features enables the network to capture a wider range of tissue characteristics. Notably, the relationship between FA and S becomes more useful when Trace is present, as regions with high Trace and low FA, such as the choroid plexus, may benefit from the additional contrast provided by sphericity. This pattern supports the view that a combination of directional, magnitude, and isotropy information is necessary for optimal parcellation.

An important observation is that although all scalar diffusion metrics are derived from the three eigenvalues of the diffusion tensor, using the raw eigenvalues directly as model input does not lead to better performance than using carefully constructed metrics such as FA, Trace, and CS. This may reflect the fact that summary metrics provide the network with features that are already aligned with biologically and anatomically meaningful properties, reducing the burden on the network to learn these relationships from scratch. In practice, this is especially important when the available training data is limited, or when trying to segment complex brain regions where contrasts are subtle and context-specific.

Further experiments reveal that a four-parameter input consisting of Trace, CS, FA, and the maximum eigenvalue (E1) achieves the highest DSC of all tested configurations. The addition of E1 likely introduces information about the dominant direction of diffusion, which can help the network distinguish between tissues with similar overall diffusivity but differing in fiber orientation. This is particularly relevant for differentiating cortical regions from subcortical structures, where directional features are more pronounced. However, performance does not continue to improve when all available diffusion parameters are included. In fact, it starts to decline, probably due to overparameterization. Providing too many input channels, especially those that are redundant or noisy, can dilute the gradients that are important for learning, reduce network generalization, and even introduce confusion for the model during training. These findings suggest a trade-off: while adding informative, non-overlapping features can help, including excessive or highly correlated parameters may hurt model performance.

Finally, our work challenges the common practice, seen in prior studies such as~\citep{li2025ddevenet, zhang2023ddparcel}, of relying solely on FA as the main input for diffusion MRI-based parcellation. While FA remains a strong and widely adopted parameter, our results make clear that it does not, by itself, yield the best possible segmentation performance. Other parameters, especially E3, as well as combinations involving Trace and CS, bring valuable, and in some cases unique, information that can improve model accuracy. Therefore, it is important for future studies to carefully consider which inputs to use, focusing on both the individual value of each parameter and the way they interact. In particular, choosing parameters that provide different, rather than overlapping, information appears to be essential for effective and generalizable parcellation pipelines.

In conclusion, our study demonstrates the importance of systematic, data-driven evaluation of input parameters in diffusion MRI-based parcellation. The selection of model inputs should be informed not just by convention or popularity in the field, but by a clear understanding of what each parameter contributes, both alone and in combination with others. Only by balancing informativeness, complementarity, and redundancy can the best possible performance be achieved in parcellating complex cortical and subcortical structures.

\subsection{Model and framework} 
\Cref{tab:SOTA} summarizes the performance of various 2.5D and 3D models tested on the HCP dataset. Notably, FastSurfer~\citep{henschel2020fastsurfer} accepts only a single input modality. Based on the results from our parameter impact study, we chose the minimum eigenvalue as the input for FastSurfer, as it exhibited the strongest contribution to parcellation performance when used individually. This choice differs from previous studies~\citep{li2025ddevenet, zhang2023ddparcel}, which employed fractional anisotropy as the input modality for FastSurfer, as part of their comparative analyses. For the models presented in~\citep{li2025ddevenet, zhang2023ddparcel}, we adhered to their original input parameter configurations, which were F+T+E1+E2+E3 and F+T+E2+E3, respectively. For all other models, we used the optimal input combination identified in our modality impact study, which was T+F+S+E1.

\begin{figure}[!t]
    \centering
    \includegraphics[width=0.95\textwidth]{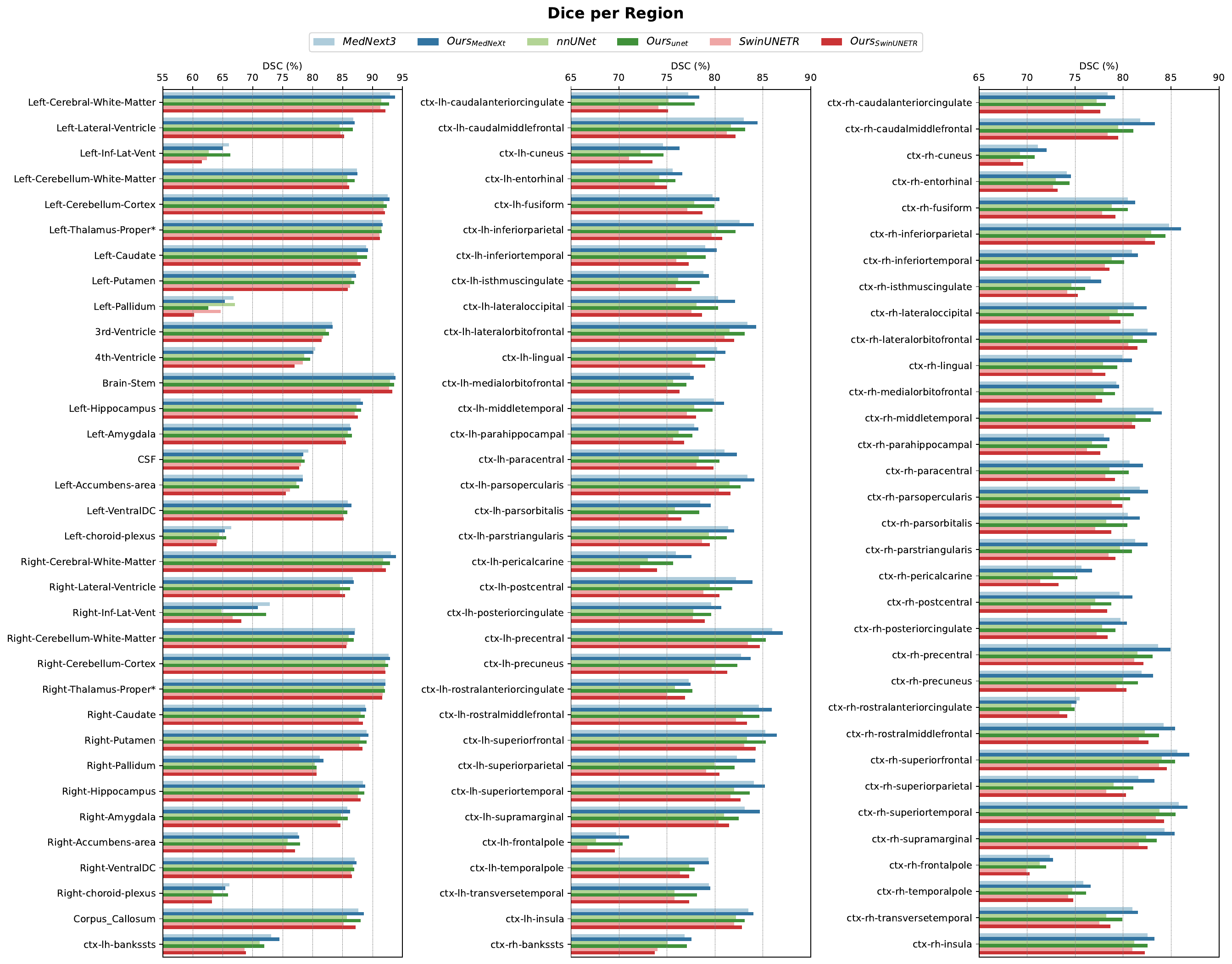} \\
    \caption{Detailed DSC for each region of the FS DK parcellation.}
    \label{fig:detailed_DSC}
\end{figure}

From the results shown in \Cref{tab:SOTA}, it is evident that 3D CNN-based models perform well in terms of parcellation accuracy. However, hybrid and transformer-based 3D models generally demonstrate lower performance compared to the CNN-based models. In particular, more efficient networks, such as SegFormer, failed to achieve accurate segmentation, resulting in poorer performance than 2D models.

Our training strategy focused on two CNN-based models, nnUNet~\citep{isensee2021nnu} and MedNeXt-M-K3\citep{roy2023mednext}, which both ranked highly in the general 3D architecture category in~\Cref{tab:SOTA}. After applying our training approach, the nnUNet model exhibited an improvement of 1.4\% in DSC and a reduction in HD95. Similarly, the MedNeXt-M-K3 model saw a 0.74\% increase in DSC, achieving a DSC greater than 82\%. These results indicate that, when trained with our strategy within the 3D framework, general-purpose 3D models outperform some models specifically designed for the DK parcellation task. Additionally, the 2.5D models, which were created specifically for this task, performed less effectively than the optimized 3D models.

\begin{figure*}[!t]
    \centering
    
    \includegraphics[width=\textwidth]{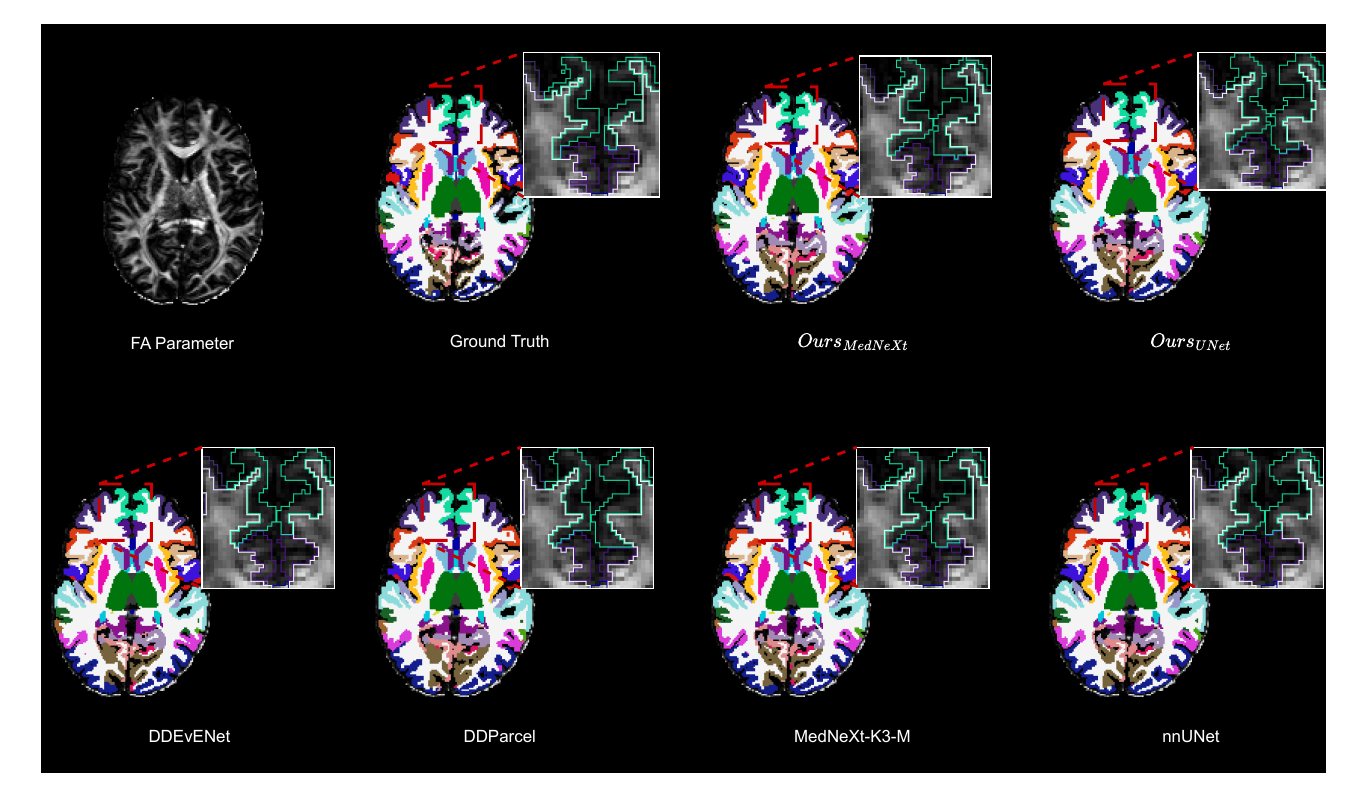}
    
    \caption{Comparison of parcellations from state-of-the-art models and our proposed framework on the HCP test set.}
    \label{fig:qualitative_results}
\end{figure*}

Interestingly, the standalone SwinUNETR model underperformed compared to nnUNet. However, when integrated into our framework, SwinUNETR showed a notable improvement of 0.87\% in DSC, surpassing the performance of the standalone nnUNet model. This finding further highlights the benefits of our focused training strategy, which allows general models to achieve high-quality results comparable to specialized models.

To further assess the statistical significance of the observed improvements, we performed a two-sample t-test to compare the DSC scores of the standalone models with those obtained when the models were embedded within our framework. The resulting p-values for nnUNet, MedNeXt-K3, and SwinUNETR were $5 \times 10^{-5}$, $1 \times 10^{-3}$, and $4 \times 10^{-6}$, respectively. These p-values provide strong evidence against the null hypothesis, supporting the conclusion that integrating these models into our framework leads to statistically significant improvements in segmentation performance. This further underscores the effectiveness of our framework in enhancing model performance across different architectures. 

A more detailed comparison of DSC scores for each region in the DK parcellation is provided in \Cref{fig:detailed_DSC}, comparing the performance of three standalone models against their performance within our dedicated framework. The results clearly demonstrate that the left and right cerebral white matter are segmented more accurately when models are used within our framework. This improvement suggests that segmenting larger regions like cerebral white matter simultaneously with smaller, more complex regions can lead to poorer segmentation outcomes due to volume disparities. When models are trained specifically to focus on these regions, as done in our framework, they can more effectively handle the segmentation task.

Additionally, \Cref{fig:qualitative_results} presents qualitative examples of the parcellations generated by the models within our framework alongside those from their standalone counterparts. These visual comparisons further demonstrate that applying a focused training strategy to general 3D models results in higher-quality dMRI parcellation outcomes.

In conclusion, our framework successfully enhances the performance of various models by providing focused, task-specific training, which improves segmentation accuracy in dMRI-based brain parcellation. This approach highlights the potential of leveraging general models with tailored training strategies to achieve state-of-the-art results in neuroimaging tasks.

\begin{table*}[!t]
\centering
\caption{RSD results of FA, MD, and Sphericity on the CNP dataset. Values are reported as the mean with the standard deviation in parentheses.}
\label{tab:RSD}
\resizebox{\textwidth}{!}{%

\begin{tabular}{c||c||c||c||c}
\bottomrule
\rowcolor{gray!10}
                 & $OURS_{unet}$  &   $OURS_{MedNeXt}$ & $OURS_{SwinUNETR}$ & T1w-reg        \\
\toprule
$RSD_{FA}$              &  0.479  (0.031)  & 0.489  (0.028) &0.485  (0.030) &0.634  (0.038)      \\  
$RSD_{MD}$              &  0.250  (0.024)  & 0.245  (0.024)& 0.242  (0.024)&0.353  (0.030)       \\  
$RSD_{S}$              &  0.179  (0.019)  & 0.180  (0.020)& 0.177  (0.019)&0.239  (0.028)       \\ 
\bottomrule
\end{tabular}


}
\end{table*}

\subsection{Model Generalization}

\Cref{tab:RSD} presents the RSD results, which compare the parcellations generated by co-registering T1-weighted images processed with FreeSurfer to the dMRI space against those produced by the nnUNet, MedNeXt-M-K3, and SwinUNETR models, all embedded within our framework, on the CNP dataset. The RSD values were calculated for three key derived parameters: FA, Trace, and CS.

Our models consistently demonstrate lower RSD values, indicating superior parcellation quality and greater homogeneity within each segmented region. These findings suggest that our framework significantly enhances the consistency and accuracy of the parcellation process across varying datasets and model configurations.

\begin{figure*}[!t]
    \centering
    
    \includegraphics[width=\textwidth]{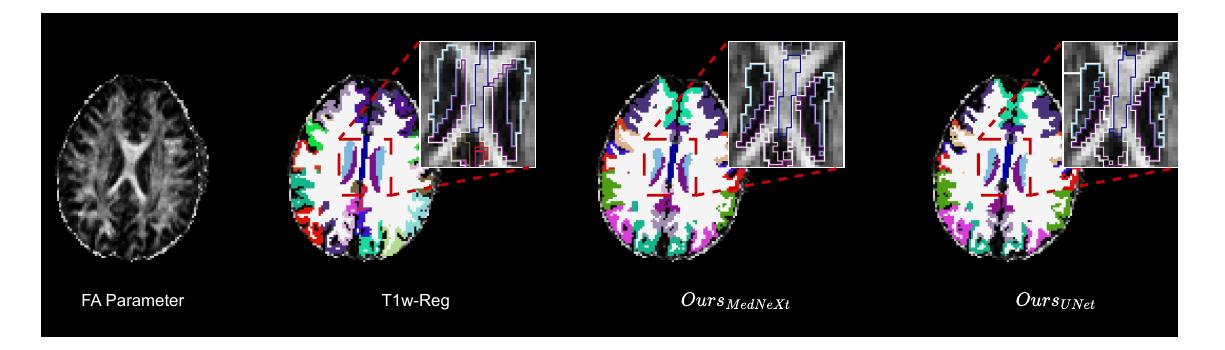}
    
    \caption{Parcellation comparison between our framework and the co-registered T1-weighted parcellation projected onto the dMRI space (T1w-reg) in the CNP dataset.}
    \label{fig:RSD_qualitative_results}
\end{figure*}

\Cref{fig:RSD_qualitative_results} presents visual comparisons of the parcellations produced by our models and those obtained through the FreeSurfer co-registration method (T1w-reg). The qualitative results show that the models integrated within our framework are more robust and exhibit better generalization across different dMRI resolutions and acquisition protocols. Notably, this improvement in parcellation quality is achieved without relying on any structural MRI data, further highlighting the effectiveness and versatility of our framework in handling diffusion MRI data independently.

\section{Conclusion}
In this work, we have introduced a novel two-stage hierarchical deep learning framework for DK brain parcellation directly from dMRI data. By employing fully three-dimensional network architectures and a coarse-to-fine segmentation strategy, our framework effectively captures the volumetric spatial context of the brain and delineates complex anatomical structures with high precision. This approach allows for accurate segmentation even in challenging brain regions, contributing to improved parcellation quality. Our comprehensive parameter study highlights the significance of selecting complementary diffusion-derived parameters, specifically fractional anisotropy, trace, sphericity, and maximum eigenvalue, to optimize segmentation performance. These findings underscore the importance of informed parameter selection to enhance model accuracy and consistency. The experimental results demonstrate that our framework outperforms current state-of-the-art methods in terms of segmentation accuracy, robustness, and generalizability. It excels across multiple datasets, exhibiting strong performance even under varying acquisition protocols and resolutions. Notably, our method eliminates the need for anatomical MRI data or complex inter-modality registration, minimizing potential sources of error and broadening its applicability across diverse research and clinical settings. In conclusion, our framework sets a new standard in dMRI-based brain parcellation. It not only enhances structural mapping accuracy but also advances neuroimaging analyses, providing a reliable tool for both neuroscience research and clinical applications. The ability to perform accurate brain parcellation directly from dMRI data, without relying on anatomical MRI, represents a significant step forward in the field.

\newpage

\section*{Data and Code Availability}

\subsection*{Data}
Our evaluation utilized two datasets: the Human Connectome Project (HCP)~\citep{van2013wu} and the Consortium for Neuropsychiatric Phenomics (CNP)~\citep{poldrack2016phenome}.

\subsection*{Code}
The implementation of our method is publicly available on \href{https://github.com/xmindflow/DKParcellationdMRI}{github.com/xmindflow/DKParcellationdMRI}.

\section*{Author Contributions}
Yousef Sadegheih: Conceptualization, methodology, software implementation, figure preparation, and manuscript writing.\\
Dorit Merhof: Supervision, funding acquisition, and manuscript review and revision.



\section*{Declaration of Competing Interests}

The authors have no competing interests to declare that are
relevant to the content of this article. 

\section*{Acknowledgements}

The authors gratefully acknowledge the computational and data resources provided by  \href{https://www.lrz.de}{the Leibniz Supercomputing Centre}. Also, the authors gratefully acknowledge the scientific support and HPC resources provided by the Erlangen National High-Performance Computing Center (NHR@FAU) of the Friedrich-Alexander-Universität Erlangen-Nürnberg (FAU) under the NHR project “b213da.” NHR funding is provided by federal and Bavarian state authorities. NHR@FAU hardware is partially funded by the German Research Foundation (DFG) – 440719683. Also, this work was supported by the German Research Foundation (Deutsche Forschungsgemeinschaft, DFG) under the grant no. 417063796.



\printbibliography
\end{document}